\documentstyle[12pt,psfig,epsf]{article}
\def\nle{\ \raise.3ex\hbox{$<$}\kern-0.8em\lower.7ex\hbox{$\sim$}\ }
\def\nge{\ \raise.3ex\hbox{$>$}\kern-0.8em\lower.7ex\hbox{$\sim$}\ }

\newcommand{\lw}[1]{\smash{\lower1.ex\hbox{#1}}}
\newcommand{\lww}[1]{\smash{\lower2.ex\hbox{#1}}}
\newcommand{\lwww}[1]{\smash{\lower3.ex\hbox{#1}}}
\newcommand{\lwwww}[1]{\smash{\lower4.ex\hbox{#1}}}
\newcommand{\lwwwww}[1]{\smash{\lower5.ex\hbox{#1}}}

\begin{document}

\baselineskip 6.25 mm
\begin{center}
{
\Large
\bf
\noindent
Collisional Stellar Dynamics, Gas Dynamics and Special Purpose Computing
}

\vspace{5mm}

Rainer Spurzem\\
(Astronomisches Rechen-Institut, Heidelberg, Germany)\\
Junichiro Makino, Toshiyuki Fukushige\\
(Dept. of Astronomy, Univ. of Tokyo, Japan)\\
Gerhard Lienhart, Andreas Kugel, Reiner M\"anner\\
(Chair No. 5, Dept. of Information Science, Univ. of Mannheim, Germany)\\
Markus Wetzstein, Andreas Burkert, Thorsten Naab \\
(Max-Planck-Inst. f\"ur Astronomie, Heidelberg, Germany)\\
\end{center}


\par\noindent

{\small

\section{Abstract}
Challenging stellar dynamical problems, such as the study of
gravothermal oscillations in star clusters, have in the past
initiated the very successful building of GRAPE special
purpose computers. It is discussed, that present day tasks
such as the formation and evolution of galactic nuclei with one or more massive
black holes and the coupled stellar and gas dynamical processes
in the formation of nuclei and star clusters, demand a new
kind of hybrid architecture, using both GRAPE and a reconfigurable
logics board called RACE. For such a system we have developed first implementations
and floating point performance studies in the case of the SPH algorithm
(smoothed particle hydrodynamics), which
will be of great advantage for SPH modelling
and also for direct $N$-body simulations using the more efficient
Ahmad-Cohen neighbour schemes.

\section{Historical Introduction}

German-Japanese cooperation in stellar dynamics dates back for at least about
two decades. One of its early highlights are the visit of D. Sugimoto as
a visiting Gau{\ss}-Professor at the university observatory in G\"ottingen,
Germany. At that collaboration, mainly with E. Bettwieser on the
German side gravothermal oscillations of globular star clusters were
detected using a gaseous model\cite{BettwieserS:84}.
The first author of this paper witnessed this as a freshman student in G\"ottingen.
Subsequently a discussion arose whether such oscillations exist in real $N$-body
systems; that question was a main motivation for the construction of GRAPE 
special purpose computers at
the University of Tokyo in Japan\cite{MakinoT:98,Sugimotoetal:90}. Their use
made it possible to demonstrate that gravothermal oscillations are indeed
present in real $N$-body systems\cite{Makino:96}. Subsequently GRAPE
became an important computing equipment for many scientists working
in stellar and galactic dynamics, also in Germany
(see informations in {\tt www.astrogrape.org}). In ongoing cooperative
projects funded by the German and Japanese science foundations the
first GRAPE computers began to work in Germany at
the University of Kiel since 1993 (later moved to Astronomisches
Rechen-Institut, Heidelberg) and at the two Max-Planck institutes (MPIA
Heidelberg, MPA Garching).
Presently, MPIA and ARI in Heidelberg, the University of Mannheim, Germany,
and the University of Tokyo cooperate on the development of hardware and
software suitable to tackle new challenges of modelling galactic nuclei,
galaxy formation and evolution, and globular clusters. In the following
a brief introduction is given into a selected sample of our astrophysical
tasks. Finally some new ideas for the use of special purpose hardware are
presented.

\section{Gravothermal Star Clusters}

Gravothermal Systems are those stellar systems in which the two-body relaxation
has played an important role in their lifetime. It can approximately be modelled
as a heat conducting gas\cite{GierszSp:94}. Astrophysical examples are
dense (globular) star clusters
and dense cusps in galactic nuclei around supermassive black holes. On the
contrary many stellar systems such as galaxies as a whole are collisionless.
Gravothermal systems are very difficult to study in direct numerical simulation,
because most pairwise gravitational interactions have to be followed with
high accuracy. The effect of gravothermal oscillations found by gas models\cite{BettwieserS:84}
could only 12 years later be seen in sufficiently large direct $N$-body simulations\cite{Makino:96}.
The search for gravothermal oscillations was one of the driving ideas
to build a fast special purpose computer called GRAPE ({\bf Gra}vity {\bf P}ip{\bf e})
\cite{MakinoT:98,Sugimotoetal:90}. The
dominant $N^2$ dependent part of the algorithm (pairwise force 
calculations) was mapped on it, using special arithmetics, pipelining and parallelisation.
Still today we have not yet reached a final understanding of how globular clusters
evolve; present studies try to improve theoretical modelling based mainly on
the Fokker-Planck approximation and to compare its results with direct $N$-body models
on rotating clusters, tidal fields and tidal shocks, a large number of primordial
binaries, the influence of stellar evolution and the final fate of its remnants. The
citation list is exemplary, but not exhaustive
\cite{Baumgardt:01,EinselSp:99,Gnedinetal:99,GierszSp:00,Hurleyetal:01,
Kimetal:01,TakaSPZ:00}.

\section{Galactic Nuclei with Black Holes}

In the course of the merger of two galaxies their
central black holes may ultimately coalesce, emitting
in the very last phase strong gravitational radiation\cite{Begelmanetal:80}.
During the early stages of the galaxy merger, the stellar
component will form a dense core through violent relaxation within a rather short
timescale. After that, the two supermassive
black holes move through the stellar component with a velocity similar
to the initial relative motion between the two galaxies. From this
moment on, both massive bodies will feel dynamical friction. This
friction leads the black holes to the newly-formed galactic center,
while the frictional force becomes more efficient with increasing
density. Through this process, the black holes must inevitably `find'
each other and form a binary system\cite{Makino:97}.

After being bound, the binary hardens (increases its binding
energy) further through dynamical
friction. At a certain binding energy dynamical friction between
each black hole and the stellar system becomes inefficient.
Close encounters and resonant three body
interactions then provide further hardening for the binary.
The most efficient process for binary hardening in this stage
are those scatterings, after which
the single stars gain very large velocities in a three body
encounter with the black holes (superelastic scatterings).
If the binary black hole centre of mass were fixed in the cluster,
this process could evacuate the
surroundings of the binary from suitable stars for further hardening, the hardening
would stall. We observe in our simulations, however, strong recoils 
of the binary centre of mass due to the superelastic scatterings, 
so the hardening does not stall\cite{Hemsendorfetal:01}. 

This is another challenging problem for our present-day software
and hardware to simulate stellar systems. New hybrid codes\cite{Hemsendorfetal:01}
including the recently
developed parallel direct $N$-body integrators\cite{Aarseth:99a,Aarseth:99b,Spurzem:99}
model a galactic nucleus containing two
massive black holes using up to 128k single particles. With the
recent GRAPE6 special purpose computer a model of three black
holes in a nucleus of 512k particles has been simulated
(Makino 2001, pers. communication). The shrinking of the
black hole binary has to be followed until a phase where
gravitational radiation sets in and a massive black hole merger
occurs.

One of the most important parameters is the final eccentricity
of the black hole binary, because the gravitational radiation
induced merger depends critically on it. We present in
Fig.~1 an example of the time evolution of the
eccentricity of the black hole binary from the work of\cite{Hemsendorfetal:01}.
A movie showing the black hole
binary forming and moving in the galactic nucleus can be found
under \par\noindent
{\tt ftp://ftp.ari.uni-heidelberg.de/pub/staff/marc/MPEG/simulation600.mpeg}.

\begin{figure}
\begin{center}
\psfig{figure=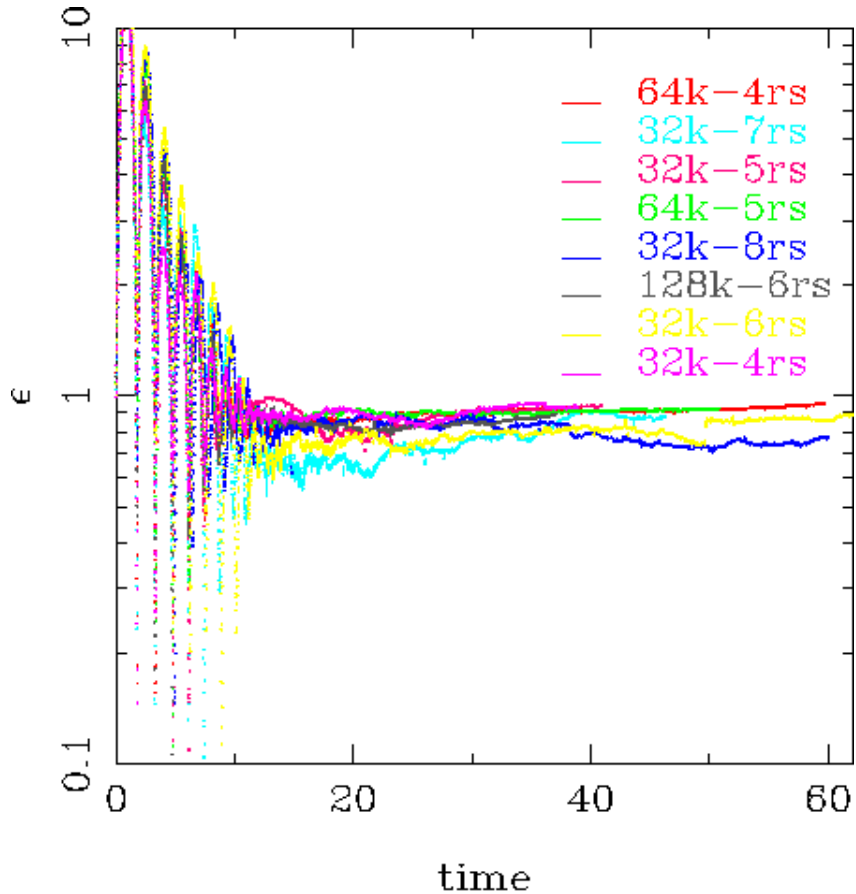,width=0.7\textwidth,angle=0}
\caption{Development of the orbit eccentricity $\varepsilon$ of the black hole binary
as a function of time in model time units (left),
for a sequence of runs with
different $N$, as indicated in the key.
Note, that if the binary is not yet bound there are formally
values of $\varepsilon >1$. Different particle numbers are indicated, and for
curves with the same $N$ the statistical particle representation
of the initial model was varied. Circular orbit is $\varepsilon=0$.}
\end{center}
\end{figure}

In the final hardening phase due to superelastic
three body encounters the eccentricity of the black hole binary remains
at a fairly large value (around 0.7), which is very interesting
because it decreases the time scale for gravitational radiation
merger of binary black holes dramatically, and thus increases
our chances to detect gravitational radiation from such events
via the planned gravitational radiation detector satellite LISA.
In a system with three black holes Makino finds even larger eccentricities
than 0.99.

\section{Gas Dynamics, Galaxies and Star Clusters}

While in certain evolutionary phases dynamics of star clusters and galactic
nuclei is predominantly determined by stellar dynamical processes
(they can be seen to first order as a system of gravitating point masses)
their formation process is necessarily linked to a complex gas physics,
including star formation, cooling, feedback processes and possibly even
exotic events such as gas production by stellar collisions or star-gas
interactions in dense star-gas systems. Again we have here for brevity
selected a few exemplary references for the interested reader which
cannot be exhaustive in any way\cite{AmaroSeoaneSp:01,Bercziketal:01,
GeyerB:01,Langbeinetal:90,NaabB:01,Nakasatoetal:01}. Many of the
cited papers use a particle approach (smoothed particle hydrodynamics\cite{Monaghan:92})
to describe the dynamics of coupled gas-star systems; it is assumed that both
stars and gas are represented by particles moving in a self-consistent
gravitational potential, but the gas particles are subject to additional
non-gravitational forces due to the gas physics.

\section{Implementation Issues - GRACE}

For astrophysical particle simulations including
self-gravity, the determination of the
full gravitational potential at each particles position is
usually the most expensive step in terms of
computational time required. This step, which is very efficiently
done by the special hardware GRAPE, requires computational
time $T = \alpha N + \beta N^2 $, where $\beta$ is a time constant
depending on hardware and on the algorithm (how many flops per individual
pairwise force calculation). The first term linearly dependant on $N$ represents
costs for advancing the particles and communication, if GRAPE is used.
The idea of
GRAPE is based on the strategy to reduce $\beta$ by hardware
for the dominant term as much as possible. The most
advanced $N$-body algorithms, however, use an Ahmad-Cohen
scheme, where the short- and intermediate range force
(neighbour or irregular force) is updated more often than the long range
(regular) force\cite{Aarseth:99a,Aarseth:99b,Spurzem:99}. 
If on average for every $\gamma$ irregular
force computations one regular force is computed the algorithm
scales as $T = \alpha N + \delta N\cdot N_n + \beta N^2/\gamma $. 
Here $N_n$ is a typical neighbour number which is small compared to $N$.
Such algorithm has been well implemented on general purpose parallel
computers\cite{Spurzem:99,SpurzemB:01}, but it does not work
efficiently on the present GRAPE special purpose hardware. 
It is interesting to note that the second scaling also holds
for high-accuracy codes using the SPH algorithm. The difference
to pure $N$-body is that $\delta$ becomes larger, since SPH forces
require of order 100 flops per pair of neighbour particles instead of only
about 20 for the gravitational neighbour force only. Also in many existing
codes the leading term is $\propto N\log N$, instead of $N^2$, because
a TREE algorithm is used for the long range force\cite{BarnesH:86}.
In the challenging applications mentioned above the high accuracy
$N^2$ algorithm will be required at least for a large subset of
particles for accuracy reasons in gravothermal
gas-star systems. Also the direct method is more advantageous for parallelisation
and the use on GRAPE computers. Whether the TREE or direct force computations
are used or not, the newly proposed hardware concept will be useful in both
cases. The development of suitable algorithms for very large particle numbers
and keeping a certain high degree of accuracy for the gravothermal systems
is subject of present and future work.

Here we propose a new hardware architecture called GRACE ({\bf GRA}PE and
R{\bf ACE}), which combines the very fast GRAPE special purpose hardware
with another hardware based on reconfigurable logics RACE (intermediate speed),
both coupled via a common PCI interface to a host workstation. 

\begin{figure}
\vspace{9cm}
\includegraphics{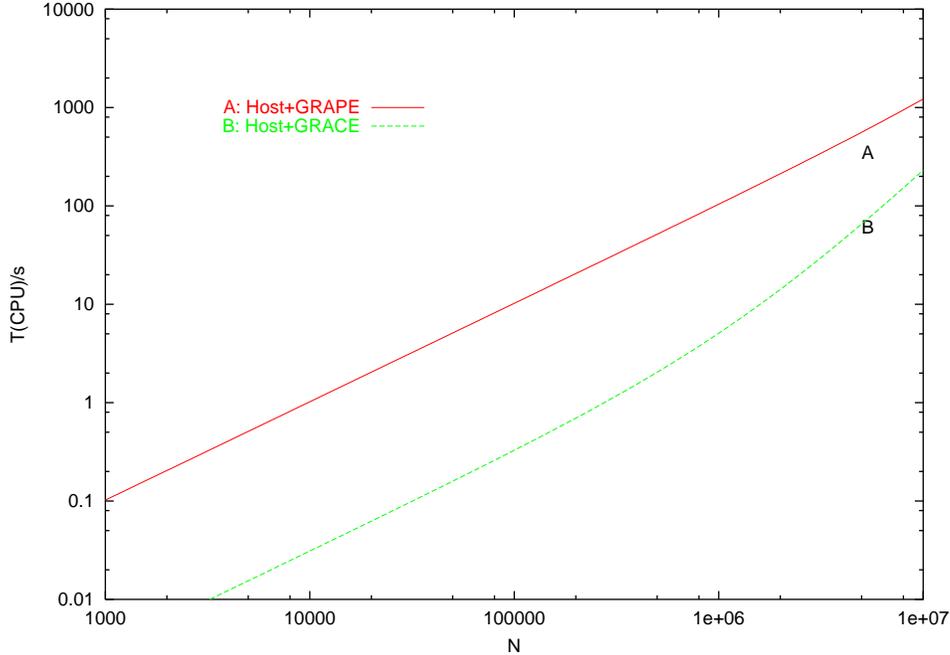}
\caption{Estimated CPU time per step for a host system with GRAPE (A),
and the combination GRACE between GRAPE and RACE (B). Assumed performance
is 50 Mflop for the host, 1 Tflop for the GRAPE, and 5 Gflop for the
RACE component, see main text.}
\end{figure}

The Fig.~2 plots a theoretical estimate to show, how
a well balanced performance of the three components GRAPE, RACE
and host workstation yields best performance. Just as an example
we use for that picture an assumed performance of 
1 Tflops for the
GRAPE subsystem, 5 Gflops for the reconfigurable RACE subsystem,
and 50 Mflop for the host workstation. In presently ongoing work
(Lienhart, G., Wetzstein, M., et al., in preparation) a full
loop of the SPH algorithm has been implemented on
the available logic resources. 
The existing implementation works on the present hardware with
roundabout 5 Gflops in floating point performance, as used in the
plot. The interested reader can find more information on the
reconfigurable RACE board under
\par\noindent
{\tt http://www-li5.ti.uni-mannheim.de/fpga/?race/}\par\noindent
and on the SPH implementation on it under \par\noindent
{\tt http://www-li5.ti.uni-mannheim.de/fpga/?astro/}.
Apart from the pure speed-up of such a combined GRACE system as
compared to a pure GRAPE system such a new hybrid architecture
has strong further potential advantages. First, it can overcome
the host bottleneck for all kinds of $N$-body-SPH applications.
Second, it can make the efficient use of GRAPE systems possible
for the more intelligent
neighbour schemes in direct $N$-body models. Third it allows
flexibility where it is required: for the short- and intermediate
range forces (softening, non-gravitational forces due to gas or
molecular dynamics). While the general idea appears very 
promising, the implementation in detail has still to be
carefully examined, in particular the way how the different
hardware components communicate and store data and neighbour
lists, for example. This work opens a path for more 
integrated GRAPE and RACE systems in the future.

 \section{Acknowledgements}
Support and warm hospitality is gratefully
acknowledged through collaborative German-Japanese grants 446 JAP 113/18/0
(1993-1998 with D. Sugimoto), 446 JAP 113/109/0 (1995-1997, with J. Makino,
K. Nomoto, S.D.M. White)
and 446 JAP 113/18/0-2 (2000-2002, with J. Makino, A. Burkert, R. M\"anner).
This research has been performed within SFB 439 ``Galaxies in the
Young Universe'' (funded by DFG) at the University of Heidelberg.
Computational Resources of the HLRS Stuttgart and
NIC J\"ulich have been used partly.

\end{document}